\documentclass[11pt,a4]{emulateapj}
\slugcomment{{\sc Accepted to ApJ:} August 22, 2011}
\usepackage{amsmath}
\usepackage{rotating}
\usepackage{hyperref}

\shortauthors{Fuentes et al. 2011}
\shorttitle{TNOs in HST, high inclination objects}

\begin{document}
\def\kms{{~\rm km\,s^{-1}}}
\def\eq#1{\rm Equation (\ref{#1})}
\def\Eq#1{\rm Eq.~\ref{#1}}
\def\Fig#1{\rm Fig.~\ref{#1}}
\def\fig#1{\rm Figure~\ref{#1}}
\def\tab#1{\rm Table~\ref{#1}}
\def\km{~\rm km\ }
\def\kmhr{~\rm km/hr\ }
\def\kpc{~\rm kpc\ }
\def\au{~\rm AU\ }
\def\yr{~\rm yr\ }
\def\Gyr{~\rm Gyr\ }
\def\hours{{\rm hours\ }}
\def\days{{~\rm days\ }}
\def\msun{M_\odot}
\def\rsun{R_\odot}
\def\lsun{L_\odot}
\def\teff{T_{\rm eff}}
\def\ms{{\rm m\,s^{-1}}}
\def\ave#1{\left<#1\right>}
\def\deg{~\rm deg}
\def\sqdeg{~\rm deg^2}
\def\aph{~\rm ''~h^{-1}}
\def\arcsec{~\rm ''}
\def\mas{~\rm mas}
\def\hr{~\rm hr}
\def\pointing{{\it pointing}}
\def\pointings{{\it pointings}}
\def\eff{{\rm eff}}

\bibliographystyle{apj}

\title{Dynamically excited outer Solar System objects in the {\it
    Hubble Space Telescope} archive\footnotemark[1]}

\author{Cesar I.\,Fuentes\altaffilmark{2},
  David E.\,Trilling\altaffilmark{2}, Matthew
  J.\,Holman\altaffilmark{3} }

\footnotetext[1]{Based on observations made with the NASA/ESA Hubble
  Space Telescope, obtained from the Data Archive at the Space
  Telescope Science Institute, which is operated by the Association of
  Universities for Research in Astronomy, Inc., under NASA contract
  NAS 5-26555. These observations are associated with program 11778.}

\altaffiltext{2}{Department of Physics and Astronomy, Northern Arizona
  University, PO Box 6010, Flagstaff, AZ 86011;
  cesar.i.fuentes@nau.edu }

\altaffiltext{3}{Harvard-Smithsonian Center for Astrophysics, 60
  Garden Street, Cambridge, MA 02138, USA}

\begin{abstract}
We present the faintest mid ecliptic latitude survey in the second
part of HST archival search for outer Solar System bodies. We report
the discovery of 28 new trans-Neptunian objects and 1 small centaur
($R\sim2$km) in the band $5^\circ-20^\circ$ off the ecliptic. The
inclination distribution of these excited objects is consistent with
the distribution derived from brighter ecliptic surveys. We suggest
that the size and inclination distribution should be estimated
consistently using suitable surveys with calibrated search algorithms
and reliable orbital information.
\end{abstract}%

\section{Introduction}\label{sec:int}

Trans-Neptunian objects (TNOs) represent the leftovers of the same
planetesimals from which the planets in the solar system formed. TNOs
offer a unique opportunity for testing theories of the growth and
collisional history of planetesimals and the dynamical evolution of
the giant planets \citep{Kenyon.2004, Morbidelli.2008}. The study of
the orbital distribution of TNOs has shown the existence of multiple
distinct dynamical populations \citep{Levison.2001,Brown.2001} with
different colors \citep{Gulbis.2010, Doressoundiram.2008} and size
distributions \citep{Bernstein.2004,Fuentes.2008}.

The population of small ($\sim$50~km) TNOs contains multiple
significant clues to understanding the formation of the Solar
System. There are more faint than bright TNOs in the Solar System,
which means that faint TNOs are a more thorough dynamical tracer, and
that some subtle clues in the dynamical distribution of TNOs may only
be revealed by studying this population.  Additionally, Pan \& Sari
showed that the size distribution of objects at this size, and in
particular the size at which the size distribution undergoes a change
in slope, records the collisional history and intrinsic strength of
the TNO population.

Because of the importance of these small --- and therefore faint, with
R$>$25 and in some cases R$>$27 --- TNOs, a great deal of effort has
been dedicated to searching for faint TNOs
\citep{Chiang.1999,Gladman.2001, Allen.2002, Bernstein.2004,
  Petit.2006, Fraser.2008, Fuentes.2008, Fraser.2009, Fuentes.2009,
  Fuentes.2010}. These surveys have been concentrated near the
ecliptic, where the sky plane density of objects is largest, since
TNOs pass through the ecliptic, regardless of inclination. However,
TNOs with inclination $i$ spend most of their time at ecliptic
latitudes $\pm i$. Few deep TNO surveys have been carried out at
ecliptic latitudes greater than a few degrees.  Consequently,
elaborate debiasing techniques have been developed \citep{Brown.2001,
  Elliot.2005, Gulbis.2010} to derive the true inclination
distribution of the TNO population. There have been no direct
measurements of the inclination distribution of faint TNOs to date.

The Hubble Space Telescope (HST) presents a unique opportunity in TNO
studies by observing across the entire sky. Observations made with HST
are deep, with a single 500~second exposure with the Advanced Camera
for Surveys (ACS) reaching $\sim$27th~magnitude, depending on the
bandpass.  The combination of these two factors implies that faint
TNOs appear serendipitously in a large fraction of all HST images,
including fields both on and off the ecliptic. The HST archive,
therefore, offers the opportunity to probe the history of the Solar
System by measuring the properties of faint TNOs at a wide range of
ecliptic latitudes.

We have developed a pipeline that harvests these serendipitous TNOs
from archival HST/ACS data \citep[hereafter F10]{Fuentes.2010}. In
F10, we searched ACS data within 5~degrees of the ecliptic and
discovered 14~TNOs, including one binary object.  Here we expand this
search to higher ecliptic latitude, 5--20~degrees. In Section~2 we
briefly summarize our field selection criteria and data processing
pipeline. Section~3 presents the objects discovered in this
mid-latitude search and their orbital parameter distribution. In
Section~4 we use this survey to test our current models of inclination
distribution.

\section{Data}\label{sec:dat}
\begin{figure*}[Ht]
  \epsscale{1.0}
  \plotone{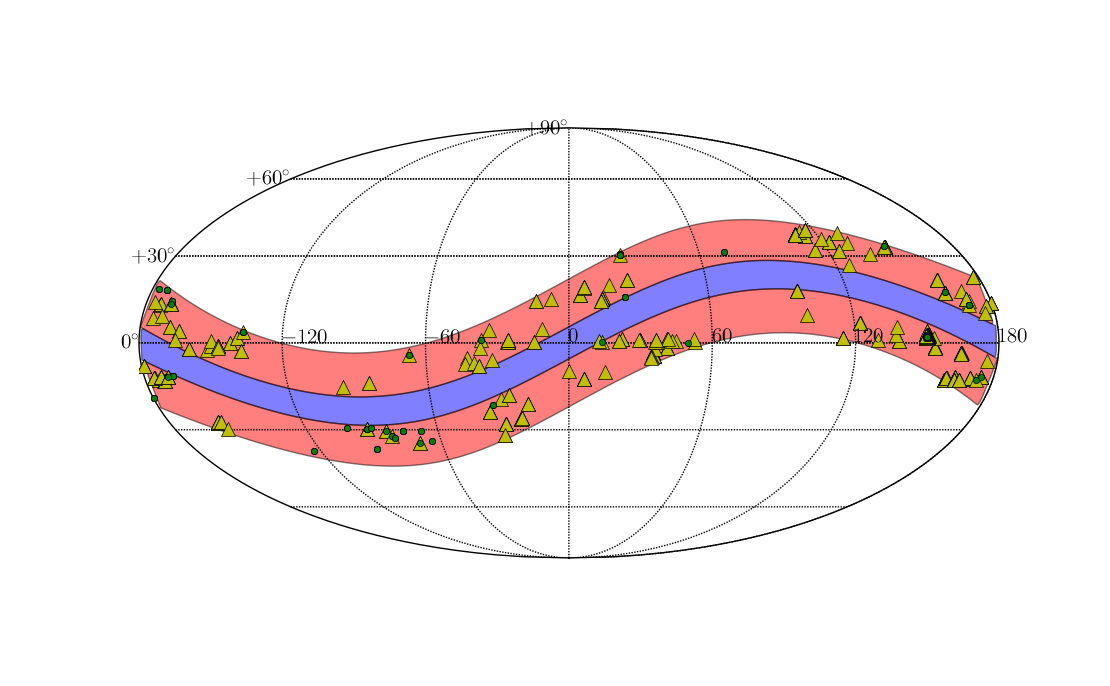}
  \caption{\label{fig:fields} Map of the sky in J2000 coordinates. The
    region we chose to select our \pointings~from is shown in red,
    that in blue is the region F10 analyzed.  The locations of all
    data considered are plotted as yellow triangles. The green circles
    are fields that were not surveyed due to problems in the
    processing or because the sky density yielded too many false
    positives. Note that there is a concentration of such fields near
    the galactic center ($-93.6,-29$) where the star density is
    highest. }
\end{figure*}

We follow the same criteria described in F10 for quick identification
of objects with TNO-like orbits in data taken with ACS/WFC. In this
survey we considered fields taken at ecliptic latitudes in the range
from 5 to 20 degrees, and for which the total exposure time within a
\pointing~was over 1,500 seconds in at least three images. The
Multimission Archive at STScI (MAST) lists data from 1,141 different
HST orbits that meet these criteria that were available as of June
4th, 2010 (See \Fig{fig:fields}).

\subsection{Pipeline}
In order to allow the implantation of a control population we
considered flat-fielded and undistorted images, as well as the filter
dependent PSF models and field distortions explained in
\cite{Anderson.2000}.

The pipeline uses the orbital solution and photometric parameters in
each image to create and implant the control population. After the
images have been cleaned of cosmic rays (CR) using a Laplacian filter
the images are ready to be searched for moving objects.

As the motion on the sky of a TNO varies greatly with its orbital
elements, we compile a set of elements that produce distinct tracks in
the images. For each set of elements in the set we produce sample PSF
tracks that are then used to find consistent sources in the
images. Finally our algorithm links sources among the images in the
set, finding all those that appear in at least 3 of them.

Then, a set of images around all plausible combinations of detections
are created to be scouted by a trained operator in order to spot
obvious contamination by chance alignment, elongated objects, poorly
subtracted CRs, etc.

\subsection{Control Population}\label{subsec:pop}
The detection and identification of a TNO in a given field of view
will depend on the characteristics of the observation and the search
algorithm. In this project we need to characterize our search method
to understand how efficient it is in finding a moving object in the
trans-Neptunian region and what variables affect the detection.

We approach this problem by creating a synthetic population that is
taken from two different orbital distributions. One is chosen to
resemble the semimajor axis distribution of the TNO population so that
objects will look like what we expect are typical TNOs. The second
distribution tries to sample the whole range of possible bound orbits
for objects at a distance between 20 and 200 AU, sampling our ability
to detect objects with unusual orbits.

For every original image 200 synthetic objects were inserted. Then the
detection process continues with no knowledge of which objects are
real and which are not. Only when real objects are identified and a
detection efficiency function is constructed is the control population
position list unveiled.

\section{Analysis}\label{sec:ana}
We run our automated search algorithm consistently for all 1141
\pointings, of which 98 failed due to too large shifts between the
images, poor signal to noise, etc. Of the remaining 1045 \pointings~
only 986 were visually inspected. A typical field would have less than
500 possible detections, so we did not inspect any field that had more
than 1,000 false positives. The 29 real objects discovered and their
best fit values for their orbital parameters are listed in
\tab{tab:obj}.

\begin{figure*}[Ht]
  \epsscale{1.0} \plotone{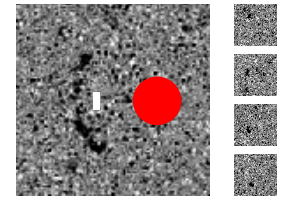}
  \caption{\label{fig:circle} Coadded image of the Centaur hst39, d=
    $12.9 \pm 0.3$ AU, over a single HST visit showing the apparent
    motion of the object. The angular size of the Earth at that
    distance is shown as a red circle. The white bar indicates the
    relative size of the off ecliptic motion of HST during the
    visit. The four individual frames are shown in sequence. }
\end{figure*}

We use images like those in \fig{fig:circle} to evaluate a $\chi
^2$ statistic when fitting for the orbital parameters of moving
objects. This is used in a Markov Chain Monte Carlo (MCMC) method to
estimate the posterior distribution function for the orbital
parameters of each found object (See \Fig{fig:d_vs_i_1} and
\Fig{fig:d_vs_e_1}).  The only constraints we considered on the
orbital parameters were a zero velocity along the line of sight and a
bound Keplerian orbit.

\begin{figure*}[Ht]
  \epsscale{1.0} \plotone{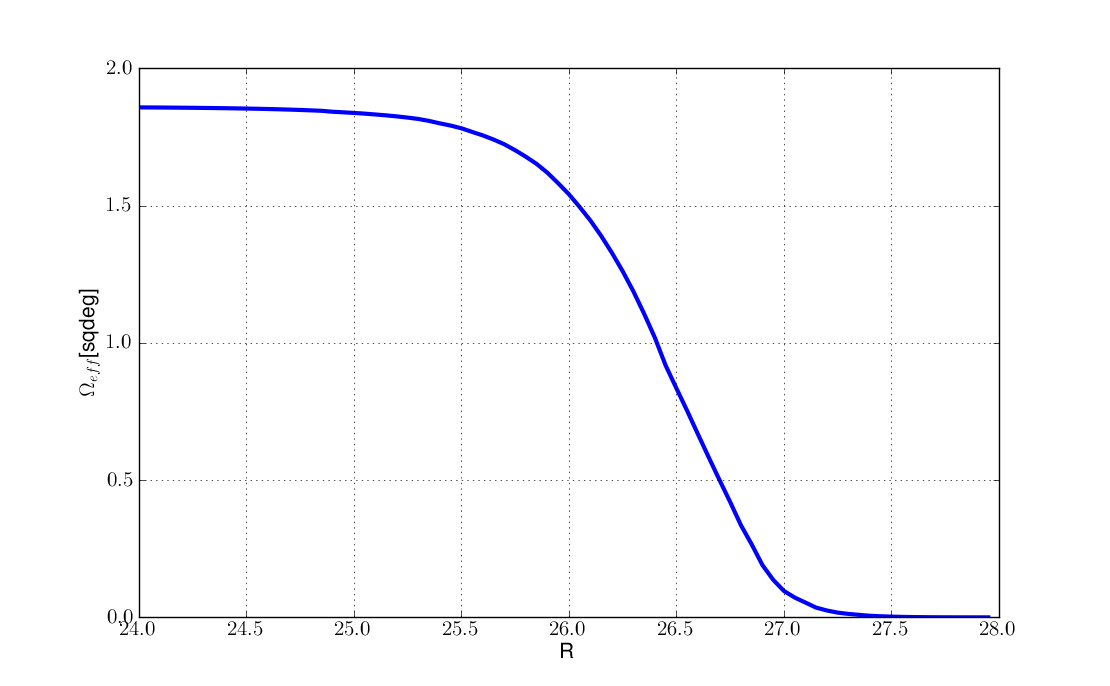}
  \caption{\label{fig:omeff} Total efficiency function for all
    \pointings~considered in this survey. It is well represented by
    the formula $\Omega_{eff}= \Omega ~{\rm erfc}(\frac{R-R_{50}}{2
      w})~ {\rm deg^2}$ where $\Omega=1.83$, $R_{50}=26.43$ and
    $w=0.29$.  }
\end{figure*}

We usually recovered half of the synthetic objects implanted, yielding
a reliable measurement of the brightness efficiency funciton for every
field. The coadded efficiency function is shown in
\Fig{fig:omeff}. This function is well represented by the formula
$\Omega_{eff}= \Omega ~{\rm erfc}(\frac{R-R_{50}}{2 w})~ {\rm deg^2}$
where $\Omega=1.83$, $R_{50}=26.44$ and $w=0.29$.

Among the moving objects discovered there is an object (hst39) whose
orbital parameters confidently constrain it to the Centaur region
($a<30$AU, Trilling et al. in preparation). This object exhibits the
least uncertainty on its distance, eccentricity and inclination, as
shown in \tab{tab:obj}. This is due to the improved resolution of the
motion of HST at the objects' distance. In \fig{fig:circle} we
show the four images taken over the course of the \pointing~coadded.

\section{Discussion}\label{sec:dis}
Of the orbital parameters for which we obtain the best constrained
quantities are the distance to the object and the inclination. The
eccentricity is also well constrained for nearer objects.

\begin{figure*}[Ht]
  \epsscale{1.}
  \plotone{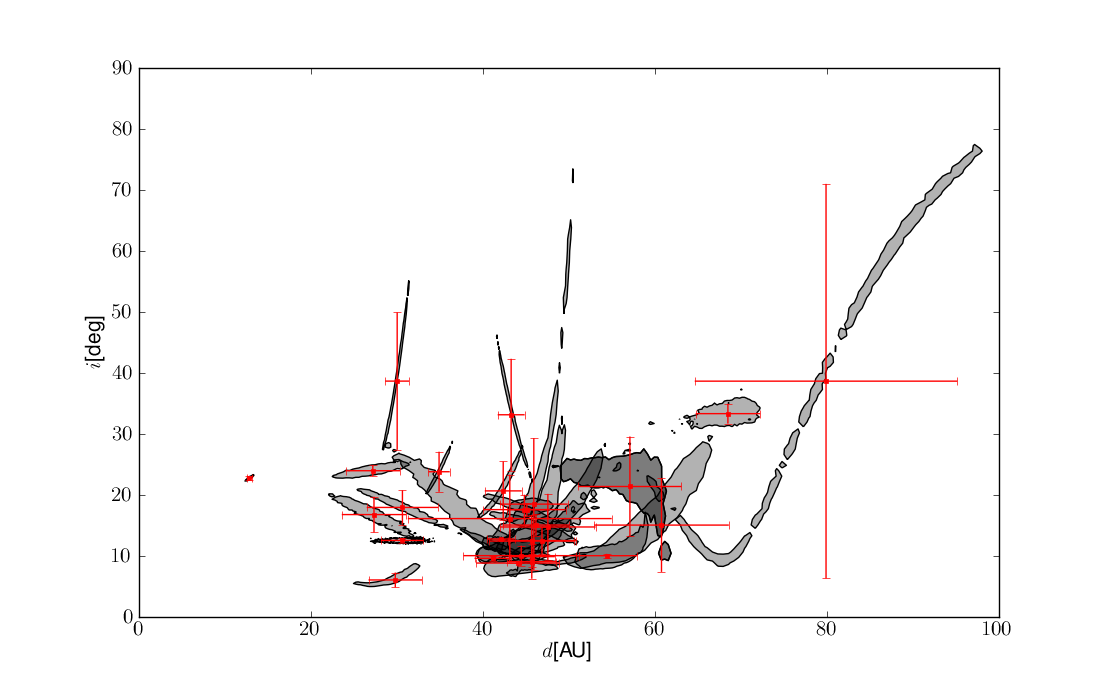}
  \caption{\label{fig:d_vs_i_1} Distance versus inclination 1$\sigma$
    probability contour for each one of the objects in
    \tab{tab:obj}. The posterior distribution function for each object
    was estimated with a MCMC as explained in the text. The confidence
    limits reported in that table are shown in red. }
\end{figure*}

\begin{figure*}[Ht]
  \epsscale{1.}
  \plotone{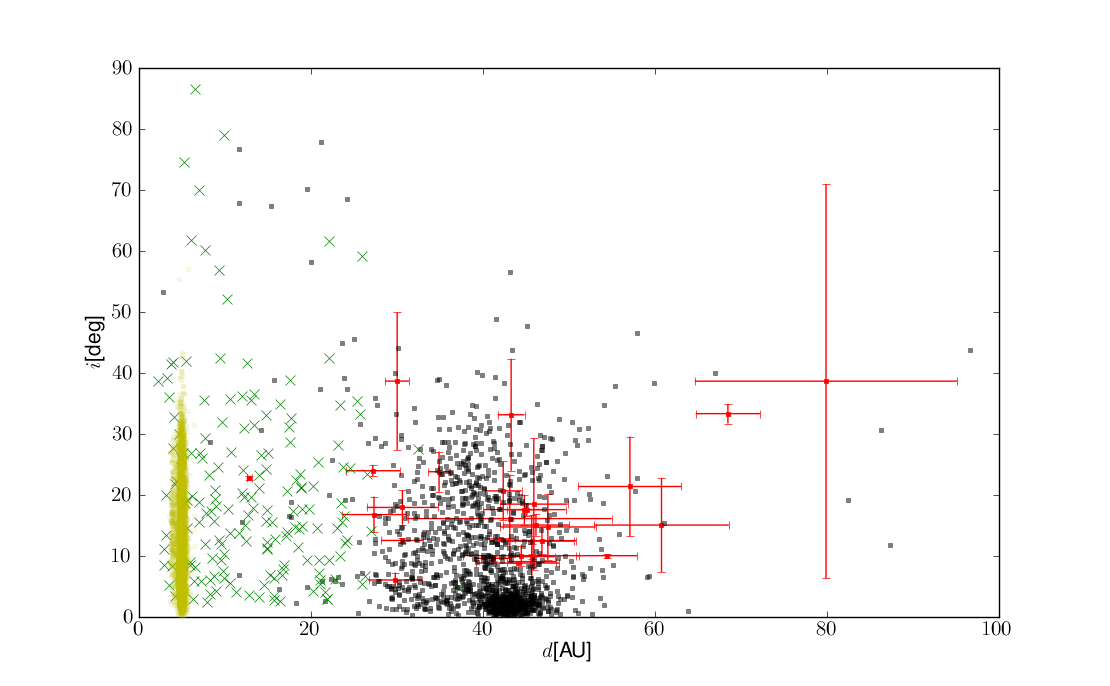}
  \caption{\label{fig:d_vs_i_2} Distance on Jan 1st, 2011 versus
    inclination. Outer solar system objects (a$>$30AU, black squares),
    centaurs (green crosses) and Jupiter trojans (yellow dots) as
    defined in the JPL lists are shown for comparison. The retrograde
    solution for hst32 is not shown. }
\end{figure*}

\begin{figure*}[Ht]
  \epsscale{1.}
  \plotone{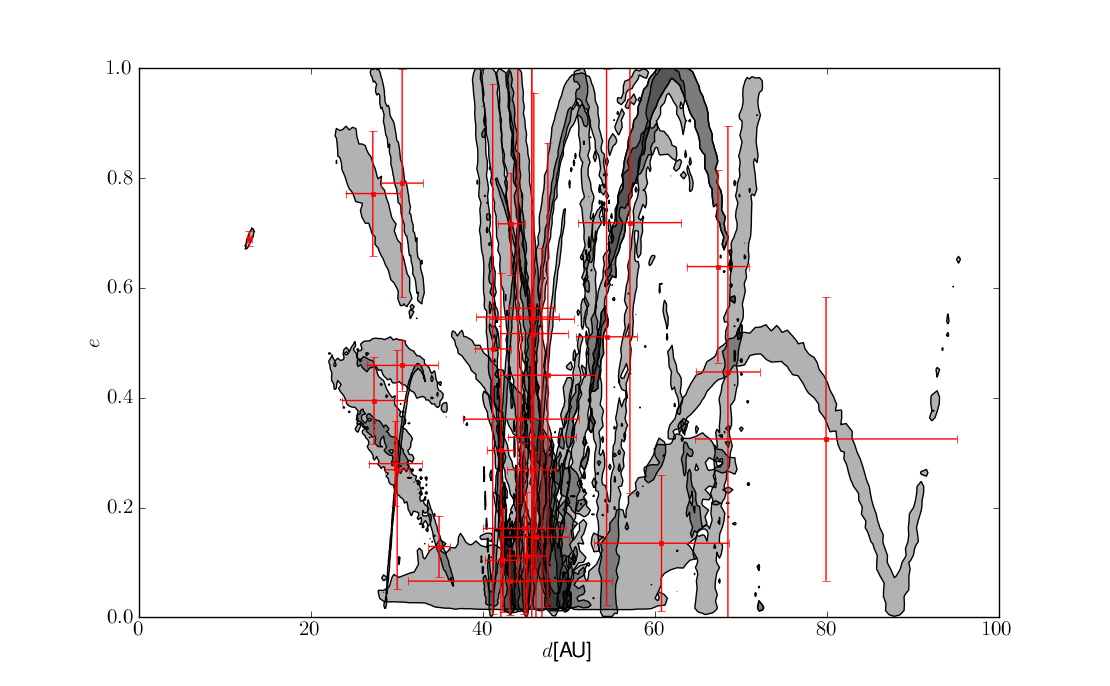}
  \caption{\label{fig:d_vs_e_1} Distance versus eccentricity 1$\sigma$
    probability contour for each one of the objects in
    \tab{tab:obj}. The confidence limits reported in that table are
    shown in red. }
\end{figure*}

\begin{figure*}[Ht]
  \epsscale{1.}
  \plotone{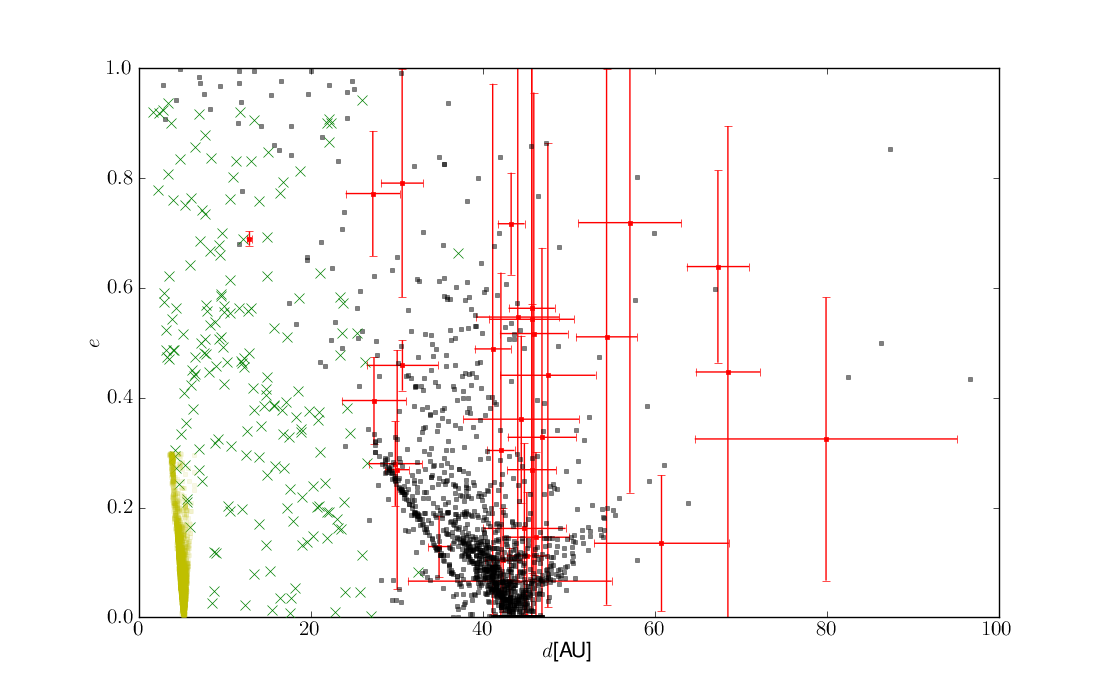}
  \caption{\label{fig:d_vs_e_2} Distance on Jan 1st, 2011 versus
    inclination. Outer solar system objects (a$>$30AU, black squares),
    centaurs (green crosses) and Jupiter trojans (yellow dots) as
    defined in the JPL lists are shown for comparison. }
\end{figure*}

We present a set of diagrams featuring the inclination
(\fig{fig:d_vs_i_1}) and eccentricity (\fig{fig:d_vs_e_1}) as a
function of distance for the objects in this survey and for the known
object in the outer solar system (\fig{fig:d_vs_i_2} and
\fig{fig:d_vs_e_2}). The information of known objects was taken from
the JPL Horizons website\footnotemark[2] and the distances were
computed for January 1st, 2011.

\footnotetext[2]{\it http://ssd.jpl.nasa.gov/}

By comparing the shape of the $1-\sigma$ confidence region in
distance, inclination and eccentricity in \fig{fig:d_vs_i_1} and
\fig{fig:d_vs_e_1} those quoted in \tab{tab:obj} we see that nominal
gaussian uncertainties are not representative of the actual range of
parameters consistent with observations. A worse situation occurs in
ground based observations, with the degrading effect from atmospheric
seeing on astrometry.

It is interesting to note that all inclination solutions have a lower
bound set by the ecliptic latitude of the field where the object was
discovered. All but the retrograde solution for hst32 are shown for
\Fig{fig:d_vs_i_1} and \Fig{fig:d_vs_i_2}. While most objects have a
relatively well constrained distance, data for hst20 is well fit by a
wide range of distances and inclinations. In \tab{tab:obj} its nominal
distance is $79.9\pm15.3$~AU which yields $H=6.9$, the lowest absolute
magnitude among our objects. Comparing the probablity distribution for
hst20's orbital parameters in \fig{fig:d_vs_i_1} and
\fig{fig:d_vs_e_1} we see that the lowest inclination corresponds to
the highest eccentricity. This solution puts hst20 much closer and in
a region with already known objects as can be seen in
\fig{fig:d_vs_i_2} and \fig{fig:d_vs_e_2}.

Some trends can be readily observed, like the high concentration of
objects at around $\sim 42$~AU in \fig{fig:d_vs_i_1} and
\fig{fig:d_vs_e_1}. The eccentricity is not well contrained except for
the few objects closer to the observer. Nevertheless, the data
constrains tight relationships between the orbital parameters.

In \fig{fig:d_vs_e_2} we see a stream of objects whose eccentricity
decreases with distance. These correspond to Plutinos in 3:2 resonance
with Neptune. There is compelling evidence for the detection of some
Plutinos in our sample by looking at \fig{fig:d_vs_e_1} where hst15,
hst18, hst25, hst27, and hst31 have distances and eccentricities
consistent with being in the 3:2 resonance.

In the inclination vs. distance diagrams we excluded the one
retrograde solution allowed by our data (See hst32 in
\tab{tab:obj}). That solution is in an area of the parameter space
completely void of known objects, while the closer prograde solution
(See \tab{tab:obj}) is well accompanied by dynamically hot TNOs.

\subsection{Inclination Distribution}
Our deep, mid-latitude survey allows us to measure the inclination
distribution of small TNOs. We assume there are only two classes of
objects in our sample of TNOs, dynamically cold and hot (those with
low and high inclinations respectively). We consider a gaussian
inclination distribution $f(i)$, and a latitudinal distribution
$L(\beta)$ like the ones described in \cite{Brown.2001} and
\cite{Gulbis.2010}.
\begin{eqnarray}\label{eq:inclidist}
f_j(i) & = & \sin(i) ~ \exp( -\frac{i^2}{2\sigma_j^2})  ~~~,\\
p_j(i | \beta) & = &  f_j(i) \frac{\cos \beta}{\sqrt{\sin^2 i - \sin^2 \beta}}  ~~~,\\
L_j(\beta) & = & K_j \int_\beta^{\pi/2}p_j(i|\beta)~di ~~~,
\end{eqnarray}
where $j$ represents the hot or cold population and $K_j$ is a
normalization constant such that $L_j(0) = 1$. The integrand $p_j(i |
\beta)$ is the expected inclination distribution from a single
observation made at a latitude $\beta$.

We consider a population of objects with distances like those of
classical TNOs and inclinations and eccentricities drawn from uniform
distributions to compute the total effective survey area. For two or
more \pointings~of the same region of the sky that were observed
within a couple of days we calculate the total effective area and
detection efficiency depending on the number of objects in our test
population that appear in multiple \pointings. The result is a set of
independent effective observations with different detection
efficiencies and observed areas. A total of 777 such observations were
identified and visually inspected from the 986 \pointings~searched.

We then compute the expected number of objects for our survey as a
function of inclination considering both the luminosity function and
inclination distribution of hot and cold TNOs. We compare this to that
of the TNOs discovered in our survey:
\begin{eqnarray}\label{eq:expected}
E(i) & = & \sum_{k=0}^{N} \sum_j
\Omega_k~\eta_k~L_j(\beta_k)~p_j(i|\beta_k) \int_{-\infty}^{\infty}
\sigma_j(R) dR ,
\end{eqnarray}
where $\sigma_j(R)$ is the number density of hot or cold objects at
the ecliptic ($\beta=0$), with the values presented in F10, and
$\eta_k$ is the detection efficiency function for pointing $k$.

\begin{figure*}[Ht]
  \epsscale{1.}
  \plotone{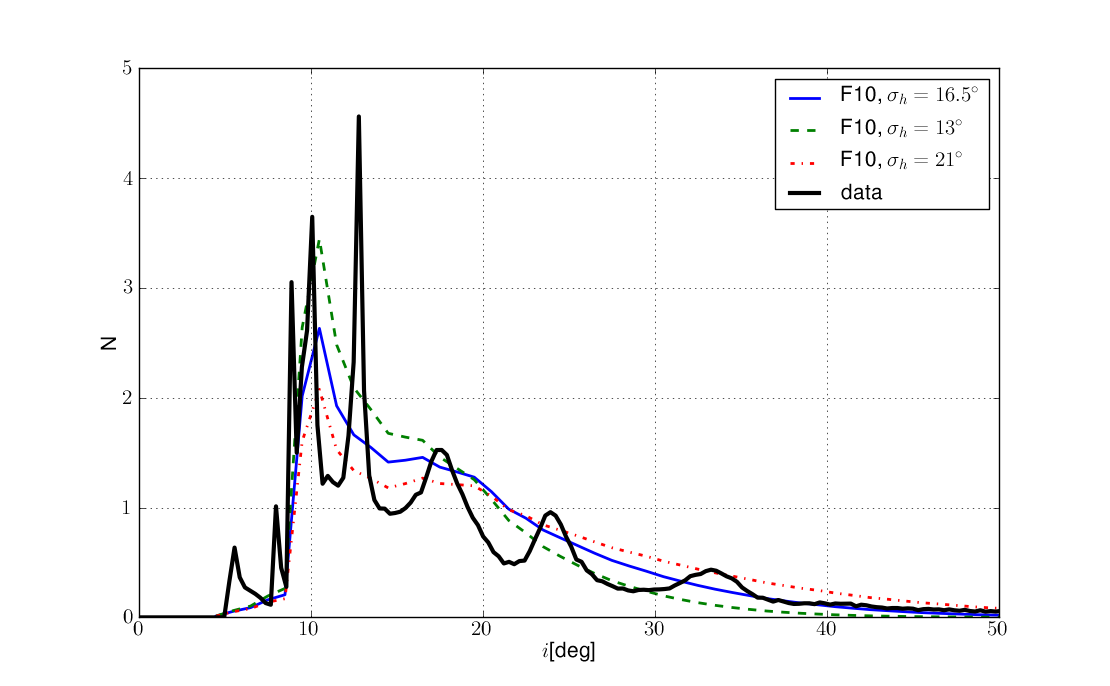}
  \caption{\label{fig:expected} The inclination probability
    distribution for all objects found in this survey is shown as a
    continuous black line. The expected probability distribution for
    different models consider the luminosity function in F10 and
    $\sigma_c=3^\circ$. The best fit for $\sigma_h=16.5^\circ$ is
    shown as a blue solid line. The limits for the $1-\sigma$
    confidence level are also shown as a dashed green line and a
    dashed-dotted red line ($\sigma_h=13^\circ,~21^\circ$
    respectively).  }
\end{figure*}

The observed inclination distribution in \fig{fig:expected} is the
addition of the inclination probability distribution for all 28 TNOs
in \tab{tab:obj}. We fit for the width of the hot population
inclination distribution $\sigma_h$ by comparing the expected and
observed probability distributions using the Kolmogorov-Smirnov
statistic. The best fit parameter is given by $\sigma_h =
{16.5}_{-3.5}^{+4.5}$~degrees. The effect of considering the cold
population is marginal, with less than one cold object expected in our
survey whether $\sigma_c = 2^\circ$ or $3^\circ$ (as computed in
\cite{Gulbis.2010} or \cite{Brown.2001} for larger TNOs). This result
indicates the TNO inclination distribution is independent of size.

There is a caveat in the the expected distribution of objects
presented in \fig{fig:expected}. The expected number of hot objects
was 50\% larger than those in our survey. To better compare expected
with observed inclination distributions we adjusted the density per
unit area for hot objects ($\Sigma_{23}$) accordingly, from 0.9 to
0.5. This only affects the number of expected hot objects, and not the
shape of the distribution.

\section{Conclusions}\label{sec:con}
We have conducted a deep search sensitive to TNOs as faint as
R$\sim$27 between ecliptic latitudes $5^\circ-20^\circ$. These are
overwhelmingly dynamically excited objects (only one cold object is
expected to appear in our sample), in contrast to surveys performed in
the ecliptic where most of the found objects are not dynamically
excited. This survey presents a unique opportunity to compare the
characteristics of dynamically excited objects to those extrapolated
from surveys carried out near the ecliptic.

The inclination distribution of TNOs has been modeled primarily with
objects discovered near the ecliptic plane \cite{Brown.2001,
  Gulbis.2010}. The double gaussian model introduced by
\cite{Brown.2001} recognized the existence of two different
populations in the classical TNOs, one dynamically excited and the
other with seemingly unperturbed near-circular orbits. Further
differences in surface properties \cite{Doressoundiram.2008} and luminosity
function \cite{Bernstein.2004, Fuentes.2010, Fraser.2010} have been
reported in the literature, hinting on an underlying difference in
size distribution and origin.

From our survey we can constrain the width of the hot inclination
distribution $\sigma_h={16.5}_{-3.5}^{+4.5}$~degrees. Changing this
parameter to the 1-$\sigma$ confidence limits gives a significantly
poorer representation of the observed distribution. The fact this
result is so close to that found by \cite{Brown.2001} is an indication
that both large and small objects have the same inclination
distribution. Comparison to the Deep Ecliptic Survey (DES) inclination
distributions in \cite{Gulbis.2010} are difficult since we do not have
enough orbital information to identify our objects in each of the DES
categories.

There are about 500 TNOs discovered in deep, well callibrated
surveys. This is enough to fit the luminosity function and inclination
distribution in a self consistent manner for the hot, cold, and
Plutino populations. The main difficulty for this is the poor
constraint on the inclination that is obtained from deep ground based
surveys. This can lead to confusion and could explain the discrepancy
in the overall number density between those discovered from the ground
(that carry most of the statistical weight in the F10 luminosity
function) and our survey.

The single centaur found at $d=13$AU in this survey (hst39) has an
estimated radius of $2$km if an albedo typical of TNOs p$\sim$7\% is
assumed \citep{Stansberry.2008}. This is one of the smallest object
found in the outskirts of the Solar System, close in size to the 500-m
sized TNO discovered by occultations \citep{Schlichting.2009}. This
discovery places a measurement of the Centaur size distribution at a
size that is comparable to the known members of the Jupiter family of
comets (JFCs, Trilling et al. in preparation).

\acknowledgments Support for program 11778 was provided by NASA
through a grant from the Space Telescope Science Institute, which is
operated by the Association of Universities for Research in Astronomy,
Inc., under NASA contract NAS 5-26555.

\clearpage

\begin{deluxetable}{llllcccccccc}
\tabletypesize{\scriptsize}
\tablecaption{\sc Found Objects}
\tablewidth{0pt}
\tablehead{
  \colhead{${\rm Name}$} &
  \colhead{${\rm MJD}$} &
  \colhead{${\rm RA}$} &
  \colhead{${\rm Dec}$} &
  \colhead{${\rm Filter}$} &
  \colhead{$\overline{m}$} &
  \colhead{$R$} &
  \colhead{$H$} &
  \colhead{${\rm opp\_ang}$} &
  \colhead{$d$\tablenotemark{a}} &
  \colhead{$i$\tablenotemark{a}} &
  \colhead{$e$\tablenotemark{a}} \\
  \colhead{} &
  \colhead{} &
  \colhead{} &
  \colhead{} &
  \colhead{} &
  \colhead{} &
  \colhead{} &
  \colhead{} &
  \colhead{$[deg]$} &
  \colhead{$[AU]$} &
  \colhead{$[deg]$} &
  \colhead{}
}
\startdata
hst15 & $53137.86081$ & 10:02:31.39 & $+$02:36:23.94 & F814W & $26.9$ & $26.2 \pm 0.3$ & $12.4$ & $99.0$ & $27.3 \pm 3.7$ & $16.9 \pm 2.9$ & $0.4 \pm 0.1$ \\
hst16 & $53138.79391$ & 10:02:24.75 & $+$02:02:28.69 & F814W & $26.1$ & $25.4 \pm 0.1$ & $9.5$ & $98.3$ & $44.8 \pm 4.8$ & $17.7 \pm 2.4$ & $0.2 \pm 0.2$ \\
hst17 & $53140.59322$ & 10:00:33.01 & $+$02:39:42.18 & F814W & $26.6$ & $25.9 \pm 0.2$ & $10.0$ & $95.9$ & $44.5 \pm 6.7$ & $10.1 \pm 1.6$ & $0.4 \pm 0.2$ \\
hst18 & $53140.65985$ & 10:00:18.85 & $+$02:42:45.20 & F814W & $26.5$ & $25.8 \pm 0.2$ & $11.6$ & $95.8$ & $30.7 \pm 4.2$ & $18.0 \pm 2.8$ & $0.5 \pm 0.0$ \\
hst19 & $53294.08604$ & 10:02:47.14 & $+$02:25:02.01 & F814W & $24.6$ & $23.9 \pm 0.4$ & $8.2$ & $51.3$ & $43.3 \pm 1.6$ & $33.2 \pm 9.2$ & $0.7 \pm 0.1$ \\
hst20 & $53336.93747$ & 10:02:40.46 & $+$02:02:44.59 & F814W & $26.0$ & $25.3 \pm 0.1$ & $6.9$ & $93.7$ & $79.9 \pm 15.3$ & $38.7 \pm 32.3$ & $0.3 \pm 0.3$ \\
hst21 & $53456.68278$ & 09:57:50.61 & $+$02:36:05.01 & F814W & $26.9$ & $26.2 \pm 0.1$ & $10.2$ & $143.3$ & $45.7 \pm 4.9$ & $12.5 \pm 4.3$ & $0.5 \pm 0.5$ \\
hst22 & $53456.68278$ & 09:58:02.27 & $+$02:36:25.01 & F814W & $26.8$ & $26.1 \pm 0.3$ & $10.5$ & $143.3$ & $41.1 \pm 2.1$ & $9.7 \pm 0.4$ & $0.5 \pm 0.5$ \\
hst23 & $53458.48194$ & 09:59:21.43 & $+$01:56:07.15 & F814W & $26.2$ & $25.5 \pm 0.7$ & $9.5$ & $142.0$ & $45.9 \pm 3.9$ & $18.6 \pm 10.8$ & $0.5 \pm 0.4$ \\
hst24 & $53599.42875$ & 22:17:22.16 & $+$00:16:47.33 & F814W & $26.3$ & $25.6 \pm 0.1$ & $8.8$ & $168.2$ & $54.4 \pm 3.6$ & $10.1 \pm 0.3$ & $0.5 \pm 0.5$ \\
hst25 & $53406.30378$ & 07:17:46.09 & $+$37:39:07.57 & F606W & $24.3$ & $23.7 \pm 0.1$ & $9.5$ & $148.6$ & $30.0 \pm 1.4$ & $38.7 \pm 11.3$ & $0.3 \pm 0.2$ \\
hst26 & $53837.23799$ & 11:19:58.11 & $+$12:59:04.06 & F555W & $24.7$ & $24.1 \pm 0.1$ & $8.1$ & $143.4$ & $45.7 \pm 2.8$ & $9.1 \pm 2.8$ & $0.3 \pm 0.3$ \\
hst27 & $53739.17629$ & 11:20:09.98 & $-$12:10:54.78 & F814W & $25.8$ & $25.1 \pm 0.1$ & $10.3$ & $107.0$ & $34.9 \pm 1.2$ & $23.9 \pm 3.3$ & $0.1 \pm 0.1$ \\
hst28 & $53871.90188$ & 21:40:09.23 & $-$23:41:43.81 & F606W & $27.3$ & $26.6 \pm 0.1$ & $9.4$ & $96.3$ & $60.8 \pm 7.8$ & $15.1 \pm 7.7$ & $0.1 \pm 0.1$ \\
hst29 & $52687.02644$ & 08:09:02.07 & $+$06:43:45.38 & F814W & $24.8$ & $24.1 \pm 0.1$ & $8.5$ & $154.3$ & $42.1 \pm 1.6$ & $12.8 \pm 0.1$ & $0.3 \pm 0.3$ \\
hst30 & $52846.01904$ & 00:26:58.72 & $+$18:57:56.91 & F814W & $25.3$ & $24.6 \pm 0.1$ & $9.0$ & $108.2$ & $42.4 \pm 2.1$ & $20.7 \pm 4.8$ & $0.1 \pm 0.1$ \\
hst31 & $53082.65966$ & 06:33:56.04 & $+$17:47:33.54 & F555W & $26.3$ & $25.6 \pm 0.1$ & $11.5$ & $99.6$ & $29.8 \pm 3.1$ & $6.1 \pm 1.2$ & $0.3 \pm 0.1$ \\
hst32\tablenotemark{a} & $53011.80363$ & 10:57:26.54 & $-$03:31:25.46 & F814W & $26.1$ & $25.5 \pm 0.3$ & $8.5$ & $119.3$ & $57.1 \pm 6.0$ & $21.5 \pm 8.2$ & $0.7 \pm 0.5$ \\
hst32\tablenotemark{a} &  &  &  &  &  &  & $7.8$ &  & $67.3 \pm 3.6$ & $153.6 \pm 1.6$ & $0.6 \pm 0.2$ \\
hst33 & $53109.33262$ & 10:00:57.23 & $+$02:31:20.99 & F814W & $26.8$ & $26.1 \pm 0.3$ & $10.1$ & $126.2$ & $45.7 \pm 2.7$ & $10.0 \pm 0.5$ & $0.6 \pm 0.5$ \\
hst34 & $53110.53235$ & 10:00:42.38 & $+$02:34:57.64 & F814W & $26.9$ & $26.2 \pm 0.1$ & $10.3$ & $125.0$ & $44.1 \pm 4.8$ & $8.9 \pm 0.1$ & $0.5 \pm 0.5$ \\
hst35 & $52936.79893$ & 09:59:43.32 & $+$01:53:12.54 & F814W & $26.4$ & $25.8 \pm 0.1$ & $8.0$ & $59.7$ & $68.5 \pm 3.7$ & $33.4 \pm 1.6$ & $0.4 \pm 0.4$ \\
hst36 & $53127.86201$ & 10:00:13.57 & $+$02:39:33.91 & F814W & $27.1$ & $26.4 \pm 0.4$ & $12.1$ & $108.0$ & $30.6 \pm 2.5$ & $12.6 \pm 0.4$ & $0.8 \pm 0.2$ \\
hst37 & $53133.79487$ & 10:00:03.34 & $+$02:37:19.99 & F814W & $27.5$ & $26.8 \pm 0.2$ & $13.0$ & $102.3$ & $27.2 \pm 3.2$ & $24.0 \pm 0.9$ & $0.8 \pm 0.1$ \\
hst38 & $53134.32813$ & 09:59:52.54 & $+$02:25:36.08 & F814W & $26.4$ & $25.7 \pm 0.1$ & $9.7$ & $101.8$ & $46.2 \pm 3.9$ & $15.1 \pm 1.8$ & $0.1 \pm 0.2$ \\
hst39 & $53129.06194$ & 10:00:02.50 & $+$02:23:52.38 & F814W & $26.7$ & $26.0 \pm 0.1$ & $15.5$ & $106.9$ & $12.9 \pm 0.3$ & $22.8 \pm 0.4$ & $0.7 \pm 0.0$ \\
hst40 & $53129.06194$ & 09:59:54.66 & $+$02:24:28.75 & F814W & $26.2$ & $25.6 \pm 0.1$ & $9.6$ & $106.9$ & $45.1 \pm 2.3$ & $17.7 \pm 1.0$ & $0.1 \pm 0.1$ \\
hst41 & $53111.53199$ & 09:58:56.65 & $+$02:14:44.58 & F814W & $26.8$ & $26.1 \pm 0.2$ & $9.9$ & $123.7$ & $47.6 \pm 5.6$ & $14.8 \pm 5.4$ & $0.4 \pm 0.4$ \\
hst42 & $53102.66744$ & 09:59:27.71 & $+$01:57:12.46 & F814W & $26.6$ & $25.9 \pm 0.1$ & $9.8$ & $132.6$ & $46.9 \pm 4.0$ & $12.6 \pm 2.3$ & $0.3 \pm 0.3$ \\
hst43 & $52845.31524$ & 14:13:16.34 & $-$01:42:04.90 & F435W & $26.5$ & $25.5 \pm 0.1$ & $9.7$ & $89.8$ & $43.1 \pm 11.8$ & $16.2 \pm 7.2$ & $0.1 \pm 0.1$ \\
\enddata \tablecomments{All objects found in this work are shown with
  their photometric and astrometric properties.  Positions given for
  the first detections. The barycentric distance $d$ and inclination
  $i$ were estimated from a MCMC with a parameterization given by the
  {\it Orbfit} code \citep{Bernstein.2000}. Though some objects were
  discovered in the same field, the epoch of the observations is
  different. The Solar System magnitude $H=V+5\log{d~\Delta}$, a
	  function of the $V$ magnitude $d$ and the distance to the observer
  $\Delta$, is computed assuming the phase angle is small and that the
  V-R color for all objects is 0.6. The conversion between HST filters
  and the Johnson system are detailed in F10. }
\tablenotetext{a}{ When prograde and retrograde solutions are possible
  we report both peaks.}
\label{tab:obj}
\end{deluxetable}
\clearpage

\end{document}